\documentclass[%
 reprint,
superscriptaddress,
 amsmath,
 amssymb,
 aps,
]{revtex4-1}

\usepackage[utf8]{inputenc}
\usepackage{subfiles}
\usepackage{graphicx}
\usepackage{hyperref}

\usepackage{amssymb}
\usepackage{amsmath}
\usepackage{braket}
\usepackage{mathrsfs}
\usepackage{amsbsy}
\usepackage{physics}
\usepackage{color, soul}
\usepackage[nonumberlist, acronym, toc]{glossaries}
\glsdisablehyper
\usepackage{algorithm}
\usepackage[noend]{algpseudocode}
\usepackage{enumitem}
\usepackage{tikz}

\usepackage{bm}% bold math

\makeatother
\DeclareMathOperator{\EX}{\mathbb{E}}% expected value
\mathchardef\mhyphen="2D
\makeatletter
     \renewcommand*\l@figure{\@dottedtocline{1}{1em}{3.2em}}
     \renewcommand*\l@table{\@dottedtocline{1}{1em}{3.2em}}    
     
\newacronym{qaoa}{QAOA}{Quantum Alternating Operator Ansatz}
\newacronym{vqe}{VQE}{Variational Quantum Eigensolver}
\newacronym{spsa}{SPSA}{Simultaneous Perturbation Stochastic Approximation}
\newacronym{rnn}{RNN}{Recurrent Neural Networks}
\newacronym{lstm}{LSTM}{Long Short Term Memory}
\begin{document}

\title{Optimizing quantum heuristics with meta-learning}

\author{Max Wilson}
\affiliation{Quantum Artificial Intelligence Lab. (QuAIL), Exploration Technology Directorate, NASA Ames Research Center, Moffett Field, CA 94035, USA}
\affiliation{Quantum Engineering CDT, Bristol University, Bristol, BS8 1TH, UK}
\affiliation{Stinger Ghaffarian Technologies Inc., Greenbelt, MD 20770, USA}
\author{Sam Stromswold}
\affiliation{Quantum Artificial Intelligence Lab. (QuAIL), Exploration Technology Directorate, NASA Ames Research Center, Moffett Field, CA 94035, USA}
\affiliation{USRA Research Institute for Advanced Computer Science (RIACS), Mountain View, CA 94043, USA}
\author{Filip Wudarski}
\affiliation{Quantum Artificial Intelligence Lab. (QuAIL), Exploration Technology Directorate, NASA Ames Research Center, Moffett Field, CA 94035, USA}
\affiliation{USRA Research Institute for Advanced Computer Science (RIACS), Mountain View, CA 94043, USA}
\affiliation{Institute of Physics, Faculty of Physics, Astronomy and Informatics, Nicolaus Copernicus University, Grudziadzka 5/7, 87–100 Toru\'n, Poland}
\author{Stuart Hadfield}
\affiliation{Quantum Artificial Intelligence Lab. (QuAIL), Exploration Technology Directorate, NASA Ames Research Center, Moffett Field, CA 94035, USA}
\affiliation{USRA Research Institute for Advanced Computer Science (RIACS), Mountain View, CA 94043, USA}
\author{Norm M. Tubman}
\affiliation{Quantum Artificial Intelligence Lab. (QuAIL), Exploration Technology Directorate, NASA Ames Research Center, Moffett Field, CA 94035, USA}
\author{Eleanor G. Rieffel}
\affiliation{Quantum Artificial Intelligence Lab. (QuAIL), Exploration Technology Directorate, NASA Ames Research Center, Moffett Field, CA 94035, USA}

% \date{\today}

\begin{abstract}
    % Quantum heuristics are cool
    Variational quantum algorithms, a class of quantum heuristics, are promising candidates for the demonstration of useful quantum computation.
    % Making them better is important
    Finding the best way to amplify the performance of these methods on hardware is an important task. Here, we evaluate the optimization of quantum heuristics with an existing class of techniques called `meta-learners'. We compare the performance of a meta-learner to Bayesian optimization, evolutionary strategies, L-BFGS-B and Nelder-Mead approaches, for two quantum heuristics (quantum alternating operator ansatz and variational quantum eigensolver), on three problems, in three simulation environments. We show that the meta-learner comes near to the global optima more frequently than all other optimizers we tested in a noisy parameter setting environment. We also find that the meta-learner is generally more resistant to noise, for example seeing a smaller reduction in performance in \textit{Noisy} and \textit{Sampling} environments and performs better on average by a `gain' metric than its closest comparable competitor L-BFGS-B. These results are an important indication that meta-learning and associated machine learning methods will be integral to the useful application of noisy near-term quantum computers.
\end{abstract}

\maketitle

\section{Introduction}
Machine learning is a powerful tool for tackling 
challenging computational problems \cite{vandal2017deepsd, libbrecht2015machine, berral2010towards}. A recent explosion in the number of machine learning applications is driven by the availability of data, improved computational resources and deep learning innovations \cite{jordan2015machine, mehta2019high, lecun2015deep}. Interestingly, machine learning has also been applied to the problem of improving machine learning models, in a field known as meta-learning \cite{vilalta2002perspective, lemke2015metalearning}. 

\begin{figure}[htb]
    \centering
    \includegraphics[width=0.45\textwidth]{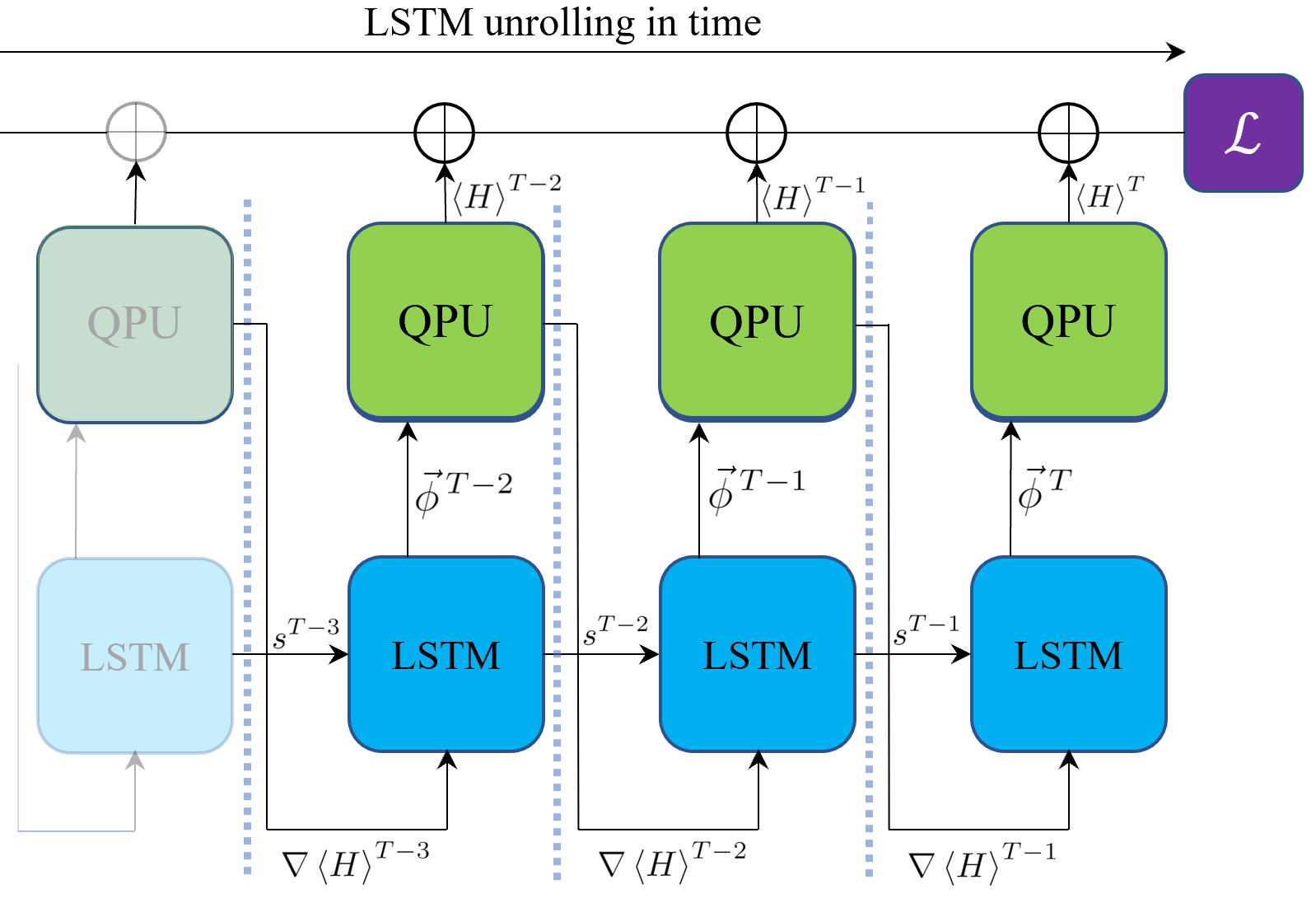}
    \caption{Meta-learner training on a Quantum Processing Unit (QPU - green). This diagram illustrates how the meta-learner used in this work can optimize the parameters of a quantum circuit (see Section~\ref{sec:setup} for a full description). Here, we outline a high level description for each time-step, such as $T-2$ (shown). A model, in our case a long short-term memory (LSTM) recurrent neural network (blue) (Section~\ref{sec:background}), takes in the gradients
    of the cost function. The LSTM outputs parameters $\vec{\phi}$ for the QPU to try at the next step. This procedure takes place over several time-steps in a process known as unrolling. The costs from each time-step are summed to compute the loss, $\mathcal{L}$ (purple), at time $T$.}
    \label{fig:unroll}
\end{figure}

In general, meta-learning is the study of models which `learn to learn'. A prominent example of a meta-learner model is one that learns how to optimize parameters of a function \cite{andrychowicz2016learning, li2016learning, ravi2016optimization,chen2017learning}.  Traditionally, this function might be a neural network \cite{andrychowicz2016learning} or a black-box \cite{chen2017learning}. Meta-learning and other new methods, including Auto-ML \cite{feurer2015efficient}, are changing the way we train, use and deploy machine learning models \cite{munkhdalai2017meta, santoro2016meta, nichol2018first}.
Here, we use a meta-learner to find good parameters for quantum heuristics, and compare that approach to other parameter optimization strategies. Figure~\ref{fig:unroll} shows an example of what the implementation of a meta-learner might look like, in the context of optimizing the parameters of a parametrized quantum circuit, illustrated as a quantum processing unit (QPU). In this work, we refer to a QPU and a quantum circuit interchangeably.

\begin{figure}[t]
    \centering
    \includegraphics[width=0.25\textwidth]{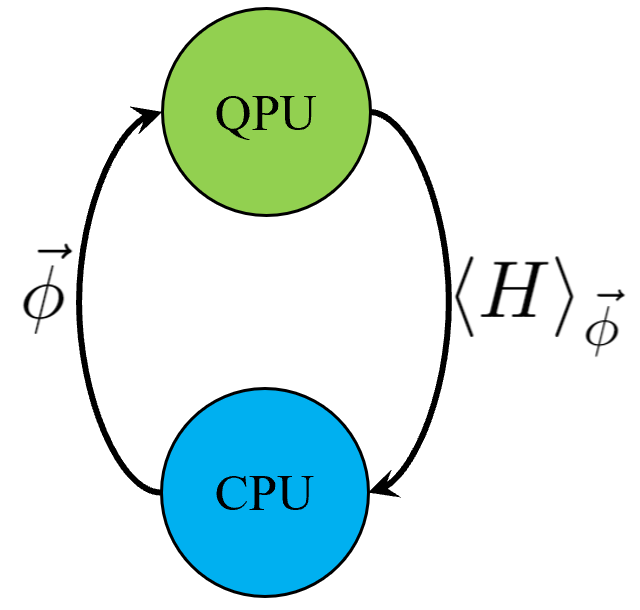}

    \caption{A single time-step of a general variational quantum algorithm, where the classical processing unit (CPU - blue) outputs parameters $\vec{\phi}$ dependent on some evaluation, in this case the expectation value $\Braket{H}$ by the quantum processing unit (QPU - green). The quantum subroutine is encoded by a  quantum circuit $U(\vec{\phi})$ (Figure~\ref{fig:circuit}) parameterized by $\vec{\phi}$, and it is responsible for generating a state $|\psi(\vec{\phi})\rangle$. This state is measured in order to extract relevant information (e.g. expectation value of a Hamiltonian). The classical subroutine suggests parameters $\vec{\phi}$ based on the values provided by a quantum computer, and sends new parameters back to the quantum device. This process is repeated until the given goal is met, i.e. convergence to a problem solution (e.g. the ground state of a Hamiltonian).}
    \label{fig:variational}
\end{figure}

Recent progress in quantum computing hardware has encouraged the development of quantum heuristic algorithms that can be simulated on near-term devices \cite{mohseni2017commercialize,preskill2018quantum}. One important heuristic approach involves a class of algorithms known as variational quantum algorithms. Variational quantum algorithms are `hybrid' quantum-classical algorithms in which a quantum circuit is run multiple times with variable parameters, and a classical outer loop is used to optimize those parameters 
(see Figure~\ref{fig:variational}).  The \gls{vqe} \cite{peruzzo2014variational}, quantum approximate optimization algorithm and its generalization \gls{qaoa} \cite{farhi2014quantum, hadfield2019quantum} are examples of algorithms that can be implemented in this variational setting. These algorithms are effective in optimization \cite{guerreschi2019qaoa, rieffel2019ans, niu2019optimizing} and simulation of quantum systems \cite{hempel2018quantum, o2016scalable, rubin2016hybrid}. The classical subroutine is an optimization of parameters, and is an important part of the algorithm both in terms of the quality of solution found and the speed at which it is found.

Techniques for the classical outer loop optimization are well-studied \cite{peruzzo2014variational, wecker2016training, wecker2015progress, guerreschi2017practical, guerreschi2019qaoa,nannicini2019performance} and several standard optimization schemes can be used. However, optimization in this context is difficult, due to technological restrictions (e.g. hardware noise), and to theoretical limitations 
such as the stochastic nature of quantum measurements \cite{knill2007optimal} or the \textit{barren plateaus} problem \cite{mcclean2018barren}.
Therefore, it is imperative to improve not only the quantum part of the hybrid algorithms, but also to provide a better and more robust framework for classical optimization.
% Finding the best way to optimize the quantum subroutine is an important part of making variational quantum algorithms useful: A good optimizer will find better solutions faster. The problem of optimization in variational quantum algorithms is well-studied \cite{peruzzo2014variational, wecker2016training, wecker2015progress, guerreschi2017practical, guerreschi2019qaoa,nannicini2019performance}, but it is still a hard problem: Evaluations of the expectation value are stochastic \cite{knill2007optimal}, hardware noise \cite{murali2019noise} and the concentration of quantum observables flattening the optimization space \cite{mcclean2018barren}, are some limiting factors. 
%Though there is a significant amount of research into enhancing machine learning with quantum computing, in a field referred to as quantum machine learning \cite{ciliberto2018quantum,benedetti2019adversarial,havlivcek2019supervised, schuld2019quantum, schuld2015introduction, grant2018hierarchical}, there is relatively less work in the area of using machine learning to improve quantum algorithms or hardware  \cite{niu2019universal, palittapongarnpim2017learning}. 
Here, we focus on the classical optimization subroutine, and suggest meta-learning as a viable tool for parameter setting in quantum circuits. Moreover, we demonstrate that these methods, in general, are resistant to noisy data, concluding that these methods may be especially useful for algorithms implemented with noisy quantum hardware.

We compare the performance of optimizers for parameter setting in quantum heuristics, specifically variational quantum algorithms. The optimization methods we compare are L-BFGS-B \cite{byrd1995limited}, Nelder-Mead \cite{nelder1965simplex}, Gaussian process regression (referred to here as Bayesian optimization) \cite{kushner1964new}, evolutionary strategies \cite{salimans2017evolution} and a \gls{lstm} recurrent neural network model \cite{hochreiter1997long} - the meta-learner. Whilst in the production of this work, we noticed similar research \cite{verdon2019learning} exploring the potential of gradient-free meta-learning techniques as initializers. Here, we use a gradient-based version of the meta-learner as a standalone optimizer (not an initializer), and a larger set of other optimizers. Though we include a diverse range of techniques, clearly, there are other optimizers that might be used, for example SPSA \cite{spall1992multivariate,spall2006theoretical,moll2018quantum, kandala2017hardware}, however our analysis focuses on those described above.

This comparison is performed in
three different simulation environments: \textit{Wave Function}, \textit{Sampling} and \textit{Noisy}. The \textit{Noisy} environment is an exact wave function simulation with parameter setting noise. The simulation environments are defined in %more 
detail in Section~\ref{sec:setup}.
% Heuristics are methods that do not necessarily have performance or runtime guarantees, but may perform well in practical applications. 
%Heuristics are methods that are not guaranteed to find optimal solutions. They are used to solve difficult real-world problems in almost every field, 
% As classical heuristics are ubiquitous across computational sciences \cite{blum2003metaheuristics}, 
%it is likely 
% it is an exciting prospect 
% that heuristics similar to those explored in this work may be the first useful application of quantum computers.

\begin{figure}[t]
    \centering
    \includegraphics[width=0.45\textwidth]{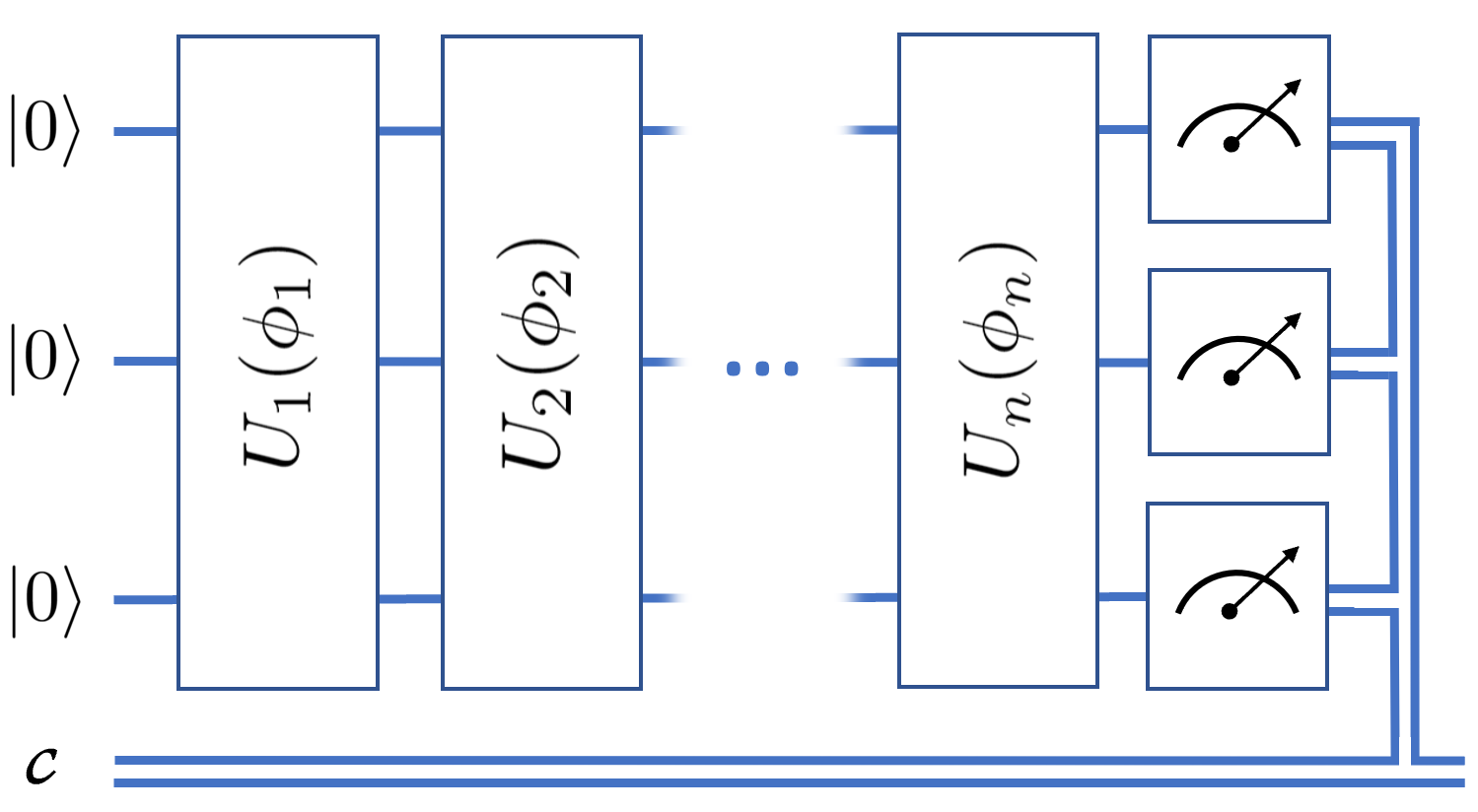}
    \caption{General parameterized quantum circuit, with arbitrary unitaries $U_j(\phi_j)$, input state $\ket{\mathbf{0}}$ and classical register $\mathbf{c}$, where $\vec{\phi} = [\phi_1,\phi_2,...,\phi_n]$ are the parameters of the circuit. Though the unitaries do not necessarily act on all qubits, we have arranged them here in `blocks', similar to the general architectures of \gls{qaoa} and \gls{vqe}, where a block of operations may be repeated many times in a circuit, with different parameters. In the case of \gls{vqe}, a block might be a series of single qubit rotations or a set of entangling gates (such as CNOT), and for \gls{qaoa}, a block might be a phase unitary encoding the cost function or a mixing unitary for searching the solution space.}
    \label{fig:circuit}
\end{figure}

The first heuristic we explore for this comparison is %the 
\gls{qaoa}~\cite{farhi2014quantum,hadfield2019quantum} for the MAX-2-SAT and Graph Bisection constraint satisfaction problems~\cite{christos1994papadimitriou}.
Second, %the 
\gls{vqe}~\cite{peruzzo2014variational} is used for estimating the ground-state of Fermi-Hubbard models \cite{hubbard1963electron}. We show that, broadly speaking, the meta-learner performs as well or better than the other optimizers, measured by a `gain' metric defined in Section~\ref{sec:sim}. Most notably, the meta-learner is observed to be more robust to noise. This is highlighted through showing the number of near-optimal solutions found in each problem by the different optimizers over all simulation environments. %as depicted in figures \ref{fig:results} and \ref{fig:bubble}. 
The takeaway of this paper is that these methods show promise, specifically the features of robustness and adaptability to hardware, and how meta-learning might be applied to noisy near-term devices. 

In Section~\ref{sec:background} we describe the background of the heuristics and optimizers. Then in Section~\ref{sec:setup} we outline the general setup including problems, the optimizers, and the simulation environments. Section~\ref{sec:sim} details the methods, including the metrics, optimizer configuration and meta-learner training. In Section \ref{sec:discussion} we discuss our results. Finally in Section \ref{sec:conc} the work is summarized and we suggest paths forward.

\section{Background}\label{sec:background}
\subsection{Quantum Alternating Operator Ans{\"a}tz}
    %The quantum approximate optimization algorithm \cite{farhi2014quantum} and its generalization the quantum alternating operator ansatz \cite{hadfield2017quantum,hadfield2019quantum} is an approach to generating families of parameterized quantum states for use in solving or approximating hard optimization problems. After creating the quantum state, a computational basis measurement is performed which returns a sample candidate solution, and the whole process is repeated while retaining the best solutions found. An important open problem is to develop strategies for determining good sets of algorithm parameters which yield good (approximate or exact) solutions with reasonably high probability. These  parameters may be determined a priori through analysis,
    %\cite{wang2018quantum},
%or searched for as part of a classical-quantum hybrid algorithm using a variational or other approach. 
%
    %Prior work on parameter setting in \gls{qaoa} includes analytic solutions for special cases  \cite{wang2018quantum}, comparison of analytical and finite difference methods \cite{guerreschi2017practical}, a method for learning a model for a good schedule \cite{wecker2016training}, and comparison of standard methods over problem classes \cite{nannicini2019performance}.

The quantum approximate optimization algorithm \cite{farhi2014quantum} and its generalization the 
quantum alternating operator ansatz \cite{hadfield2019quantum} (\gls{qaoa}) form families of parameterized quantum circuits
for generating %(approximate) 
solutions to combinatorial optimization problems. After initializing a suitable quantum state, a \gls{qaoa} circuit consists of a fixed number~$p$ %$p=1,2,\dots$ 
blocks (see Figure~\ref{fig:circuit}), where each block
is composed of a phase unitary generated from the %combinatorial problem 
cost function 
we seek to optimize, followed by a mixing unitary.
%and consisting of $z$ rotations of angle $\gamma$. This is followed by a mixer stage consisting of $x$ rotations. 
The phase unitary 
%is typically compiled as 
typically yields 
a sequence of multiqubit Pauli-$Z$ rotations each with phase angle $\gamma$. 
In the original proposal of Farhi et al.~\cite{farhi2014quantum}, the mixing unitary is a Pauli-$X$ rotation of angle $\beta$ on each qubit. However, 
extending the protocol to more general encodings and problem constraints naturally leads to 
a variety of more sophisticated families of  mixing operators~\cite{hadfield2017quantum,hadfield2019quantum}.
%may be derived   
%it is possible to %improve upon this original framework extend
%by adapting the mixer to the appropriate problem representation \cite{hadfield2017quantum,hadfield2019quantum}. 
%This combination of unitaries is repeated $p$ times before 
At the end of the circuit a measurement is performed in the computational (Pauli-$Z$) basis to return a candidate problem solution.

An important open %problem 
research area is to develop strategies for determining good sets of algorithm parameters (i.e. the $\gamma$ and $\beta$ values for each block) which yield good (approximate or exact) solutions with 
%reasonably high 
nonnegligible probability.
These  parameters may be determined a priori through analysis,
%\cite{wang2018quantum},
or searched for as part of a classical-quantum hybrid algorithm using a variational or other approach. Prior work on parameter setting in \gls{qaoa} includes analytic solutions for special cases  \cite{wang2018quantum}, comparison of analytical and finite difference methods \cite{guerreschi2017practical}, a method for learning a model for a good schedule \cite{wecker2016training}, and comparison of standard approaches over problem classes~\cite{nannicini2019performance}.

We evaluate parameter setting strategies for \gls{qaoa} for
%the well-known 
MAX-2-SAT and Graph Bisection, %problems, 
both NP-hard combinatorial optimization problems \cite{christos1994papadimitriou,ausiello2012complexity}. We use standard`\cite{farhi2014quantum} and generalized~ \cite{hadfield2019quantum} QAOA methods, respectively. 
The latter problem mapping is of particular interest as it %attacks an important optimization problem using 
utilizes an advanced family of QAOA mixing operators from \cite{hadfield2019quantum} %related to the XY model from physics,
%that have been
 that has recently been demonstrated to give advantages over the standard mixer \cite{wang2019xy}.  

\subsection{Variational Quantum Eigensolver}
The \gls{vqe} \cite{peruzzo2014variational} is a hybrid optimization scheme built on the variational principle. It aims to estimate the ground state energy of a problem Hamiltonian through iterative improvements of a trial wave function. The trial wave function is prepared as a quantum state using a parameterized quantum circuit, and 
the expectation value of the Hamiltonian with respect to this state is measured. This energy value is then passed to a classical device, which uses optimization techniques (SPSA, BFGS, etc.) to update the parameters. The process is repeated for a fixed number of iterations, or until a given accuracy achieved.

% The \gls{vqe} is an optimization approach that uses the variational principle in order to optimize wave functions of interest.  Approaches that make use of the variational principle are widely used, and is the basis for variational quantum Monte Carlo approaches on classical computers.  Regardless of the computing architecture being used, the main idea of using the variational principle is to optimize a wave function ansatz that is parameterized by variables that can be optimized. The ansatz does not necessarily allow for arbitrary wave function forms, however they are generally chosen to such that they can accommodate a wave function that has large overlap with a state of interest, in many cases this state is the ground state of a Hamiltonian.  

% Examples of various ansatz that are used on classical computers include matrix product states, coupled cluster wave functions, and truncated CI expansions.   On a quantum computer it has been proposed various forms of tensor networks and cluster expansions (Unitary coupled cluster for example), should be used as they are generally hard to simulate classically and they have potential to be improvements on the wave function ansatz we have classically. 

The initial demonstration of \gls{vqe} used Nelder-Mead, a standard derivative-free approach, for parameter setting after observing that gradient descent methods did not converge \cite{peruzzo2014variational}. Since then, examples in the literature include the use of \gls{spsa} in \cite{moll2018quantum}, where the authors argue simultaneous perturbation methods might be particularly useful for fermionic problems, but classical problems (such as MaxCut) may favor more standard techniques (i.e. gradient descent). Other routines used include COBLYA, L-BFGS-B, Nelder-Mead and Powell in \cite{romero2018strategies}. Finally, in \cite{moseley2018bayesian} the authors explore the use of Bayesian optimization for parameter setting in \gls{vqe}.

\subsection{Meta-learning}

%%% Meta learning
Meta-learning is the study of how to design machine learning models to learn fast, well and with few training examples \cite{bengio1990learning}. One specific case is a model, referred to here as a meta-learner \cite{ravi2016optimization}, which learns how to optimize other models. A model is a parameterized function. Meta-learners are not limited to training machine learning models; they can be trained to optimize general functions \cite{chen2017learning}. In the specific area of using models to optimize other models, early research explored Guided Policy Search \cite{li2016learning}, which has been superceded by \gls{lstm}s \cite{andrychowicz2016learning, chen2017learning, bello2017neural, wichrowska2017learned}. An \gls{lstm} is a recurrent neural network, developed to mitigate vanishing or exploding gradients prominent in other recurrent neural network architectures \cite{bengio1994learning, hochreiter1998vanishing}. It consists of a cell state, a hidden state, and gates, and all three together are called an LSTM cell. At each time-step, changes are made to the cell state dependent on the hidden state, the gates (which are models) and the data input to the LSTM cell. The hidden state is changed dependent on the gates and the input. The cell state and hidden state are then passed to the LSTM cell at the next time-step. A full treatment of an LSTM is given in Reference~\cite{hochreiter1997long}. An LSTM is good for learning long-term (over many time-steps) dependencies, like those in optimization.

Meta-learners have been used for fast general optimization of models with few training examples \cite{ravi2016optimization}: Given random initial parameters we seek to achieve a fast convergence to `good' (defined by some metric) general parameters. This same problem feature appears for \gls{qaoa}, where good parameters may follow some common distribution across problems \cite{wecker2016training}. A meta-learner could be used to find general good parameters, and fine-tuning left to some other optimizer \cite{verdon2017quantum}, though this approach was not explored here.

\section{Setup}\label{sec:setup}

\subsection{Simulation Environments}

We compare optimization methods in `\textit{Wave Function}', `\textit{Sampling}' and `\textit{Noisy}' simulation environments. The \textit{Wave Function} case is an exact wave function simulation. For \textit{Sampling}, the simulation emulates sampling from a hardware-implemented quantum circuit, where the variance of the expectation value evaluations is dependent on the number of samples taken from the device. In these experiments, we set the number of shots (samples from the device) to 1024. 

\begin{figure}[b]
    \centering
    \includegraphics[width=0.42\textwidth]{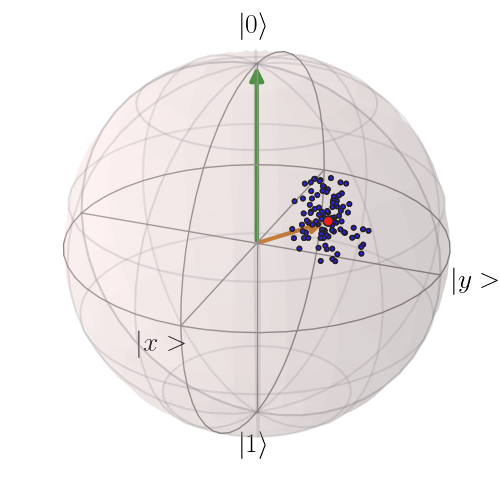}
    \caption{Rotation of initial state $\ket{0}$ (green) by rotation operator $R_Z(\pi/4)R_Y(\pi/3)R_Z(0)$ to new state (orange arrow, red point). When noise of $\sigma=0.1$ is applied to the parameter setting we see a distribution of final states (blue) over 100 trials.}
    \label{fig:bloch}
\end{figure}

\begin{figure}
    \centering
    \includegraphics[width=0.48\textwidth]{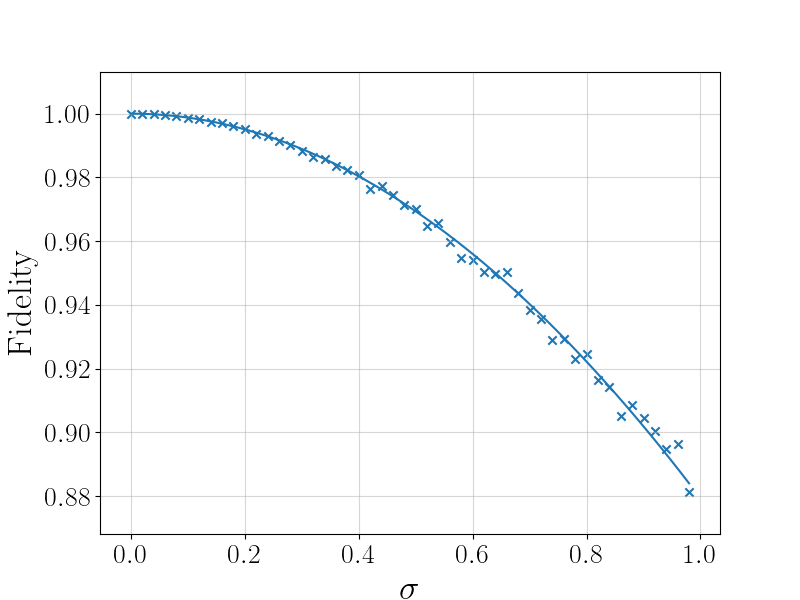}
    \caption{Effective single qubit rotation gate fidelity plotted as a function of the noise on input parameters. Parameters are sampled from a normal distribution with standard deviation $\sigma$ and centered on the target input value.}
    \label{fig:noise}
\end{figure}

Lastly, in the \textit{Noisy} case we have modelled parameter setting noise in an exact wave function simulation. We assume exact, up to numerical precision, computation of the expectation value (via some theoretical quantum computer which can compute the expectation value of a Hamiltonian given a state up to arbitrary precision). Then, for each single-qubit rotation gate, we added normally distributed, standard deviation $\sigma = 0.1$, noise to the parameters at each optimization step. In order to determine $\sigma$, we evaluated the relationship between the fidelity of an arbitrary rotation (composed of three single-qubit Pauli rotation gates $R_Z(\alpha)R_Y(\beta)R_Z(\gamma)$), % Figure \ref{fig:bloch}) 
around the Bloch sphere and parameter noise; see Figure~\ref{fig:bloch}. Assuming industry standard single qubit gate rotations of $~99\%$ \cite{krantz2019quantum}, a value of $\sigma=0.1$ is approximated, see Figure~\ref{fig:noise}. 
All simulations were performed with Rigetti Forest \cite{forest} simulators.


\subsection{Optimizers}

\subsubsection{Local optimizers}

Nelder-Mead and L-BFGS-B are gradient-free and gradient-based approaches, respectively, which are  standard local optimizers~\cite{guerreschi2017practical, wecker2016training, wecker2015progress, nannicini2019performance}.
Local optimizers have a notion of location in the solution space. They search for candidate solutions from this location. They are usually fast, and are susceptible to finding local minima.  L-BFGS-B is a local optimizer and has access to the gradients. Out of all optimizers chosen it is the closest to the meta-learner in terms of information available to the optimizer and computational burden (i.e. the cost of computing the gradients). Nelder-Mead was chosen as it appears throughout the literature \cite{peruzzo2014variational,guerreschi2017practical, verdon2017quantum, romero2018strategies} and provides a widely recognized benchmark.

\subsubsection{Bayesian Optimization}

Global optimizers are designed to search for a global optima, and are generally more computationally intensive.  An important class of global black-box optimizers we consider are Bayesian optimizers. 

Bayesian optimization, also known as Gaussian process regression, involves computing updates to a posterior probability distribution over candidate functions, given a prior distribution and training examples \cite{shahriari2015taking}. The training examples are function input-output pairs. The runtime of Bayesian optimization scales as~$O(N^3)$ where~$N$ is the number of training points. It is useful for finding a global optima with the minimum number of steps \cite{kushner1964new}. 

\subsection{Evolutionary Strategies}

Evolutionary strategies are %also a 
another class of global black-box optimization techniques: A population of candidate solutions (individuals) are maintained, which are evaluated based on some cost function. Genetic algorithms and evolutionary strategies have been used for decades. More recent work has shown these techniques to be competitive in problems of reinforcement learning~\cite{vidnerova2017evolution,salimans2017evolution}.

All implementations of evolutionary strategies are population-based optimizers. In the initial iteration, the process amounts to a random search. In each iteration, solutions with lower costs are more likely to be selected as parents (though all solutions have a nonzero probability of selection). Different methods for selecting parents exist, but we used binary tournament selection, in which two pairs of individuals are selected, and the individual with the lowest cost from each pair is chosen to be a parent.

In more precise terms, parents are the candidate solutions selected to participate in crossover. Crossover takes two parent solutions and produces two children solutions by randomly exchanging the bitstring defining the first parent with the second. Each child replaces its parent in the population of candidate solutions. The process is repeated, so costs for each child are evaluated, and these children are used as parents for the next iteration~\cite{beasley1993evolution}. In our case, the bitstring is divided into $n$ subsections, where $n$ is the number of parameters passed to the quantum heuristic. Each subsection is converted to an integer using Gray encoding and then interpolated into a real value in the range $[-\pi /2, \pi /2]$. Gray codes are used as they avoid the Hamming walls found in more standard binary encodings \cite{charbonneau2002evolution}.

%Genetic algorithms are a particular type of evolutionary algorithm and work by encoding solutions into a string of bits. Given a cost function $f: \mathbf{R}^n\mapsto\mathbf{R}$, we consider a bitstring (chromosome) of length $nm$ where each component of the vector $\mathbf{R}^n$ is encoded in $m$ bits. Each parameter $\theta_i$ may assume values in the range $[a_i, b_i]$ (in this case $a_i=-b_i=\pi /2$). Bitstrings are converted to integers using Gray coding; this is standard practice as it reduces the impact of Hamming walls \cite{charbonneau2002evolution}. Integers are then linearly interpolated into the valid $[a_i, b_i]$ range.

%If a given parameter may assume values in the interval $[a_i, b_i]$, a binary representation is converted into a real value by the function
%\begin{equation}
%f(\mathbf{x})=G(\mathbf{x})\frac{b_i-a_i}{2^{m}-1} + a_i
%\end{equation}
%where $G(\mathbf{x})$ is the integer Gray decoding of the bitstring $\mathbf{x}$.  
It is the bitstrings that are operated on by the genetic algorithm. When two individuals are selected to reproduce, a random crossover point, $b_c$ is selected with probability $P_c$. Two children are generated, one with bits left of $b_c$ from the first parent and bits to to the right of $b_c$ originating from the second parent. The other child is given the opposite arrangement. Intuitively, if $b_c$ is in the region of the bitstring allocated to parameter $\phi_k$, the first child will have angles identical to the first parent before $\phi_k$ and angles identical to the second parent after $\phi_k$. Again, the second child has the opposite arrangement. The effect on parameter $\phi_k$ is more difficult to describe. Finally, after crossover is complete, each bit in each child's bitstring (chromosome) is then flipped (mutated) with probability $P_m$. Mutation is useful for letting the algorithm explore candidate solutions that may not be accessible through crossover alone.

Evolutionary strategies are highly parallelizable, robust and relatively inexpensive \cite{salimans2017evolution}. Both Bayesian optimization and evolutionary strategies are good candidates for optimizing quantum heuristics and are used here.

\subsubsection{Meta-learning on quantum circuits}
The meta-learner used in this work is an \gls{lstm}, shown
unrolled in time in Figure~\ref{fig:unroll}. Unrolling is the process of iteratively updating the inputs, $x$, cell state and hidden state, referred to together as $s$, of the \gls{lstm}. Inputs to the model were the gradients of the cost function w.r.t. the parameters, preprocessed by methods outlined in the original work \cite{andrychowicz2016learning}. At each time-step they are
\begin{equation}
    x^t =
    \begin{cases}
      (\frac{\log(\abs{\nabla \Braket{H}^t})}{r},\mathrm{sign}(\nabla \Braket{H}^t)) & \text{if}\ \abs{\nabla \Braket{H}^t}\geq e^{-r} \\
      (-1,\exp(r)\nabla \Braket{H}^t), & \text{otherwise}
    \end{cases}
    \label{eq:preprocessing}
\end{equation}
where $r$ is a scaling parameter, here set to $10$, following standard practice \cite{andrychowicz2016learning, ravi2016optimization}. The terms  $\nabla \Braket{H}^t$ are the gradients of the expectation value of the Hamiltonian at time-step $t$, with respect to the parameters $\vec{\phi}^t$. This preprocessing handles potentially exponentially-large gradient values 
whilst maintaining sign information. Explicitly, the meta-learner used here is a local optimizer. At some point $\vec{\phi}^t$ in the parameter-space, where $t$ is the time-step of the optimization, the gradients $x^t$ are computed and passed to the LSTM as input. The LSTM outputs an update $\Delta \vec{\phi}^t$, and the new point in the parameter space is given by $\vec{\phi}^{t+1} = \vec{\phi}^t + \Delta \vec{\phi}^t$. 
It is possible to use these models for derivative-free optimization \cite{chen2017learning}, however, given that the gradient evaluations can be efficiently performed on a quantum computer, scaling linearly with the number of gates, and that the optimizers usually perform better with access to gradients, we use architectures here that exploit this information. In Reference \cite{mcclean2018barren} the authors show that the gradients of the cost function of parameterized quantum circuits may be exponentially small as a function of the number of qubits, the result of a phenomena called the concentration of quantum observables.
In cases where this concentration is an issue, there may be strategies to mitigate this effect \cite{grant2019initialization}, though it is not an issue in the small problem sizes used here.

Though only one model (a set of weights and biases) defines the meta-learner, it was applied in a `\textit{coordinatewise}' way:
For each parameter a different cell state and hidden state of the \gls{lstm} are maintained throughout the optimization. Notably, this means that the size of the meta-learning model is only indirectly dependent on the number of parameters in the problem. We used a gradient-based approach, exploiting the parameter-shift rule \cite{schuld2019evaluating} for computing the gradients of the loss function with respect to the parameters. These were used at both training and test time.

All model training requires some loss function. We chose the summed losses, 
\begin{equation}
    \mathcal{L}(\omega) = \EX_f \big[ \sum_{t=0}^T \omega_t f(\phi_{t}) \big],
    \label{eq:loss}
\end{equation}
where $\EX_f$ is the expectation over all training instances $f$ and T is a time-horizon (the number of steps the \gls{lstm} is unrolled before losses from the time-steps $t<T$ are accumulated, backpropagated, and the model parameters updated). The hyperparameters $\omega_t$ are included, though are set to $\omega_t = 1$ %\,\, \forall \,\, t$
for all $t$ 
in these training runs. This can be adjusted to weigh finding optimal solutions later in the optimization more favourably, a practice for balancing exploitation and exploration. In situations where exploration is more important, other loss functions can be used, such as the expected improvement or observed improvement \cite{chen2017learning}. However, in this instance we chose a loss function to rapidly converge, meaning fewer calls to the QPU. This has the effect of converging to local minima in some cases, though we found that this loss function performed better than the other gradient-based optimizer (L-BFGS-B) for these problems.

\subsection{Problems}

\subsubsection{Fermi-Hubbard Model}

Hubbard Hamiltonians have a simple form, as follows:

\begin{align} \label{eqn:hubbard}
    H = &- t \sum_{\langle i,j \rangle}\sum_{\sigma=\{\uparrow,\downarrow\}}(a^\dagger_{i, \sigma} a_{j, \sigma} + a^\dagger_{j, \sigma} a_{i, \sigma}) \\ 
    &+ U \sum_{i} a^\dagger_{i, \uparrow} a_{i, \uparrow} a^\dagger_{i, \downarrow} a_{i, \downarrow}- \mu \sum_i \sum_{\sigma=\{\uparrow,\downarrow\}} a^\dagger_{i, \sigma} a_{i, \sigma},\nonumber
\end{align}

where $a_{i,\sigma}^\dag, a_{i,\sigma}$ are creation and annihilation operators, respectively, of a particle at site $i$ with spin $\sigma$. In this model there is a hopping term $t$, a many body interaction term $U$ and an onsite chemical potential term~$\mu$.  This model gained importance as being a possible candidate Hamiltonian to describe superconductivity in cuprate materials.  However, recent numerical studies have shown that there are some significant differences between the model and what is seen in experiments, such as the periodicity of charged stripes that the model supports~\cite{Simons2015,hubbard1,hubbard3}.  However, the model is quite interesting itself, with many different phases of interest.   The model is also quite difficult to solve, especially when going to large lattice sizes and large values of $U/t$.  This has lead to many studies and much method development on classical computers, and is still widely researched today.

\begin{figure}[h]
    \centering
    \includegraphics[width = 0.25\textwidth]{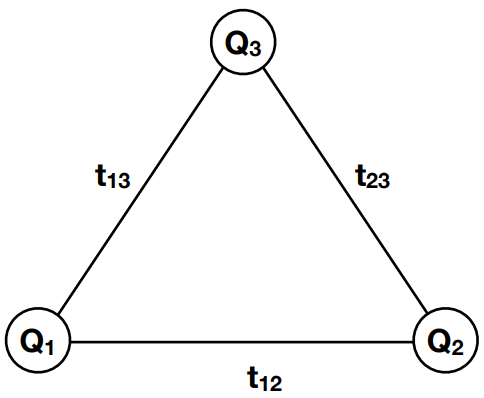}
    \caption{Sketch of a spinless three-qubit Fermi-Hubbard model that is used for the VQE optimization. Coupling strengths are not necessarily equal and take values from $[-2,2]$. }
    \label{fig:FH}
\end{figure}

For %the 
\gls{vqe} we look for the ground-state of the simplified spinless three-site Fermi-Hubbard model with unequal coupling strengths $t_{ij}\in[-2,2]$ and $U=\mu=0$, Figure~\ref{fig:FH}. The Hamiltonian of this model can be mapped through the Jordan-Wigner transformation \cite{Jordan1928} to the qubit Hamiltonian

\begin{align}
    H_{FH} = \frac{1}{2}\Big(t_{12}X_1X_2 &+ t_{12}Y_1Y_2 +t_{23} X_2X_3 + t_{23}Y_2 Y_3 \nonumber \\
    &+ t_{13}X_1Z_2X_3 + t_{13}Y_1Z_2 Y_3\Big).
    \label{eq:FermionicTriangle}
\end{align}

Based on the results of \cite{Woitzik2018,Woitzik2019}, we use a circuit composed of 3 blocks. Each block consists of three single qubit rotations $R_Z(\alpha)R_Y(\beta)R_Z(\gamma)$ applied to all qubits, followed by entangling CNOT gates acting on qubits (1,2) and (2,3), where the first entry is the control qubit and the second is the target. 

\subsubsection{MAX-2-SAT}
Given a Boolean formula on $n$ variables in conjunctive normal form (i.e. the AND of a number of disjunctive two-variable OR clauses), MAX-SAT is the NP-hard problem of determining the maximum number of clauses which may be simultaneously satisfied. %When each clause contains at most $k\geq 2$ variables the problem is called MAX-$k$-SAT.
The best classical efficient algorithm known achieves only a constant factor approximation in the worst case, as deciding whether a solution exists that obtains better than a particular constant factor is NP-complete~\cite{christos1994papadimitriou}.
For MAX-2-SAT, where each clause consists of two literals, 
the number of satisfied clauses can be expressed as 
\begin{equation}
 %C(x) = 
 %Note %C(x) not same as cost Ham C below which is negated
 C = \sum_{(i,j)\in E} \Tilde{x}_i \lor \Tilde{x}_j
\end{equation} 
where $\Tilde{x}_i$ in each clause represents the binary variable~$x_i$ or its negation, and $E$ is the set of clauses.
We use an $n$-qubit problem encoding where the $j$th qubit logical states $\ket{0}_j,\ket{1}_j$ encode the possible values of each~$x_j$.
Transforming to Ising spin variables~\cite{hadfield2018representation} and substituting with Pauli-$Z$ matrices
leads to the cost Hamiltonian %in terms of Pauli $\sigma}_z^$ matrices as 
\begin{equation}
    \widehat{C} =  \sum_{(i,j)\in E} \frac{1}{4} (1 \pm \hat{\sigma}_z^{(i)})(1\pm\hat{\sigma}_z^{(j)}) 
\end{equation}
which is minimized when the number of satisfied clauses is maximized. The sign factors 
$+1$ or $-1$ in $\widehat{C}$ correspond to whether each clause contains $x_i$ or its negation, respectively. Note that $C$ and $\widehat{C}$ are \textit{not} equivalent; $C$ gives a maximisation problem, while $\widehat{C}$ gives a minimization problem, with the same set of solutions.

For our QAOA implementation of MAX-2-SAT we use the original~\cite{farhi2014quantum} initial state $\ket{s}=\tfrac1{\sqrt{2^n}}\sum_x \ket{x}$, phase operator $U_P(\widehat{C},\gamma)=\exp(-i\gamma \widehat{C})$, and mixing operator $U_M(\beta)=\exp(-i\beta \sum_{j=1}^n \sigma^{(j)}_x)$. The example instances we consider below have $n=8$ qubits, 8 clauses, and \gls{qaoa} circuit depth $p=3$. 

\subsubsection{Graph Bisection}
Given a graph with an even number of nodes,
the Graph Bisection problem is to partition the nodes into two sets of equal size such that the number of edges across the two sets is minimized. The best classical efficient algorithm known for this problem provably yields only a $\log$-factor worst-case  approximation ratio \cite{krauthgamer2006polylogarithmic}. 
Both this problem and its maximization variant are NP-hard \cite{christos1994papadimitriou}.

For an $n$-node graph with edge set $E$ we encode the possible node partitions with $n$ binary variables, where~$x_j$ encodes the placement of the $j$th vertex. In this encoding, from the problem constraints the set of feasible solutions is encoded by strings $x$ of Hamming weight $n/2$. 
The cost function to minimize can be expressed as 
\begin{equation}
    C = \sum_{(i,j) \in E} \mathrm{XOR(x_i, x_j)}
\end{equation}
under the condition $\sum^n_{j=1} x_j = n/2$. Transforming again to Ising variables gives the cost Hamiltonian 
\begin{equation}
    \widehat{C} = \frac12 \sum_{(i,j) \in E} (1 - \hat{\sigma}_{z}^{(i)}\hat{\sigma}_{z}^{(j)}).
\end{equation}
A mapping to QAOA for this problem was given in \cite[App. A.3.2]{hadfield2019quantum} from which we derive our construction. We again encode possible partitions $x$ with the  $n$-qubit computational basis states $\ket{x}$. For each problem instance we uniformly at random select a string $y$ of Hamming weight $n/2$ and use the feasible initial state $\ket{y}$. The phase operator $U_P(\widehat{C},\gamma)=\exp(-i\gamma \widehat{C})$ is constructed in the usual way from the cost Hamiltonian.
For the mixing operator we employ a special case of the $XY$-mixer proposed in \cite{hadfield2019quantum}. 
%given by a sequence of $n$ partial mixers 
%$U_M(\beta)= U_n(\beta) \dots %U_2(\beta)U_1(\beta)$.
This class of mixers affect state transitions only between states of the same Hamming weight, which will importantly restrict the quantum state evolution to the feasible subspace.  
For each node $j=1,\dots,n$, we define the $XY$ partial mixer 
$$U_j(\beta)=\exp\left(-i\beta \left( \hat{\sigma}_X^{(j)} \hat{\sigma}_X^{(j+1)}+
\hat{\sigma}_Y^{(j)} \hat{\sigma}_Y^{(j+1)} \right) \right)$$
with $\sigma^{(n+1)}:=\sigma^{(1)}$. We define the overall mixer to be the ordered product 
$U_M(\beta)= U_n(\beta) \dots U_2(\beta)U_1(\beta)$. Observe that as each partial mixer preserves feasibility, so does $U_M(\beta)$, and so \gls{qaoa} will only output feasible solution samples. We consider problem instances with $n=8$ qubits, 8 edges, and \gls{qaoa} circuit depth $p=3$.

\section{Methods}\label{sec:sim}

\subsection{Metrics}
Here, we outline two metrics used to evaluate and compare the optimizers. The first metric used is the gain, $\mathcal{G}$, to the minimum,
\begin{equation}
    \mathcal{G} = \EX_f \Big[ \frac{f_F-f_I}{f_\mathrm{min}-f_I}\Big]
    \label{eq:G}
\end{equation}
where $\EX_f$ is the expectation value over all instances $f$, $f_F$ is the converged cost of the optimizer, $f_I$ is the initial cost (determined by the initial parameters) and $f_\mathrm{min}$ is the ground-state energy. $f_\mathrm{min}$ was determined by evaluating all possible solutions in the cases of MAX-2-SAT and Graph Bisection, and by exact diagonalization of the Hamiltonian for finding the ground-state of the Fermi-Hubbard model.
This number is the expectation over instances $f$ of the `gain' to the global minimum from the initialized parameters. In the case of local optimizers (meta-learner, L-BFGS-B, Nelder-Mead) we initialized to the same parameters. The metric outlines the average progress to the global minimum from an initialization.
Secondly, the quality of the final solution was also evaluated by a distance to global minima metric, $\mathcal{D}$, 
\begin{equation}
    \mathcal{D} = \frac{\abs{f_{\mathrm{min}} - f_F}}{\abs{f_\mathrm{min}-f_\mathrm{max}}} * 100
    \label{eq:D}
\end{equation}
where $f_\mathrm{max}$ is the maximum possible energy. This metric gives a sense of the closeness to the global minima, as a percentage of the extent. 

\subsection{Configuring Optimizers}

We evaluated the optimizers on 20 problems from 5 random initializations each, to increase the probability of reaching the ground-state by all optimizers. The initializations were kept the same between the local optimizers (L-BFGF-B, Nelder-Mead and meta-learner). Global optimizers used 5 different random initializations for each problem (evolutionary strategies and Bayesian optimization). L-BFGS-B and Nelder-Mead were implemented using Scipy \cite{scipy}, where the gradients for L-BFGS-B were computed by analytic means and quantum circuit simulation. These optimizers were left to converge at later iterations and not terminated at 100. Bayesian optimization was implemented with GPyOpt \cite{gpyopt2016}, and terminated at convergence.
We implemented and configured the evolutionary strategies methods in-house. For all tests, a small population size of 20 was used to limit the number of calls to the simulator (sizes on the order of 100 are typical and may improve performance). Both MAX-2-SAT and Graph Bisection problems with QAOA used $m=60$ bits to represent parameters. VQE simulations had more parameters to optimize, so $m=297$ bits were used for these problems. All tests used a probability of crossover of $P_c=0.9$, and a probability of mutation of $P_m=0.01$. 

% We did not parallelize implementation of the genetic algorithm: Limited access to quantum computation resources will prevent this in the near-term.

\begin{figure*}[htb]
    \centering
    \includegraphics[width=\textwidth]{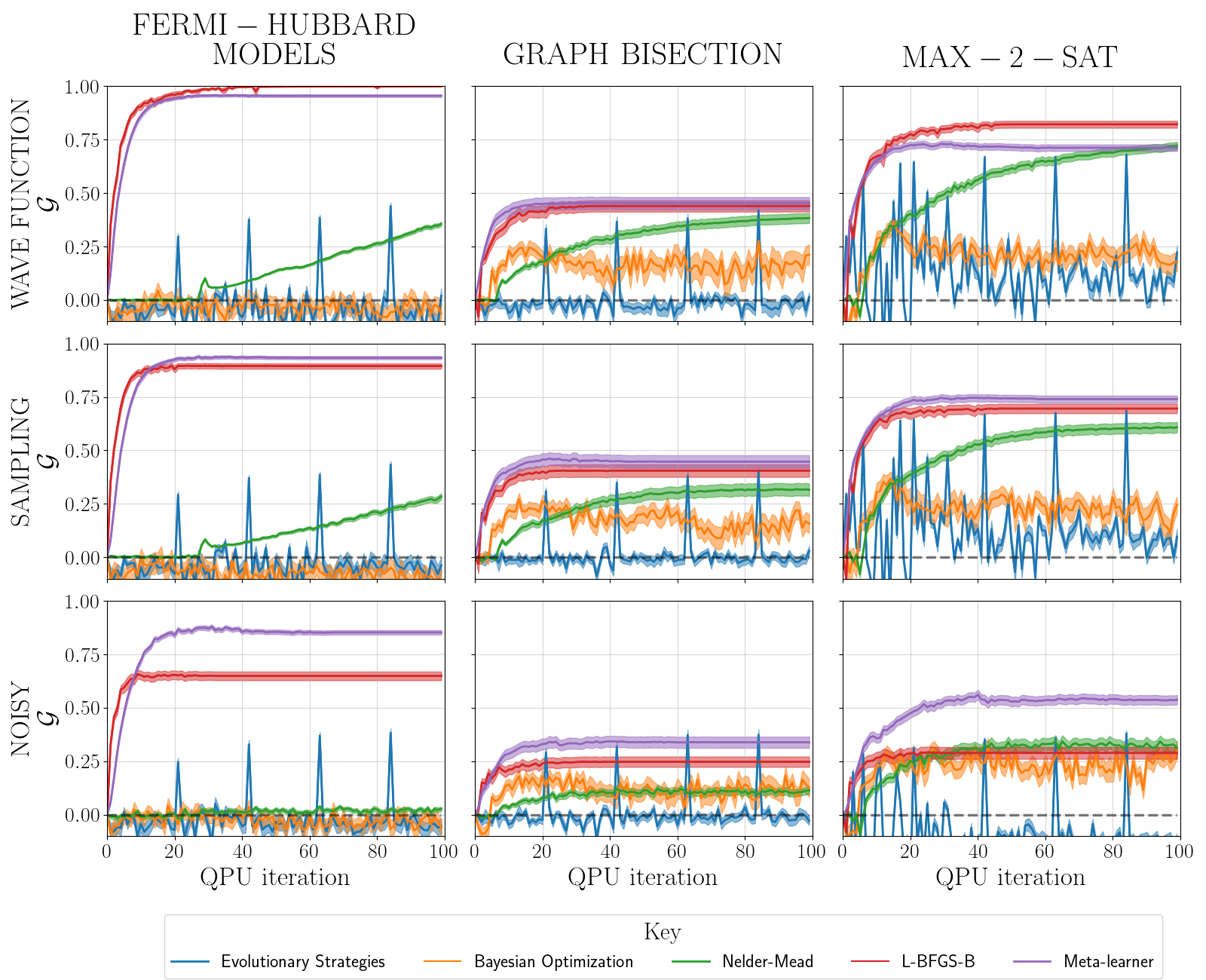}
    \caption{Left to right columns: Fermi-Hubbard models, Graph Bisection and MAX-2-SAT problems. Top to bottom rows: \textit{Wave Function}, \textit{Sampling} and \textit{Noisy} simulations, defined in Section \ref{sec:setup}. Optimizers: Evolutionary strategies (blue), Bayesian optimization (orange), Nelder-Mead (green), L-BFGS-B (red), meta-learner (purple). x-axis: Shared within a column, QPU iteration is number of times $\Braket{H}$ has been evaluated. y-axis: Shared within a row, $\mathcal{G}$, the gain, is the value computed by Equation~(\ref{eq:G}), and represents the average progress toward the minimum from the initial evaluation of $\Braket{H}$. We recognise that this comparison is \textit{not} apples to apples: L-BFGS-B and the meta-learner have access to the gradient, and make numerous calls to auxiliary quantum circuits (simulated in the same environment as the expectation value evaluation circuits) to compute the gradients. The number of calls to evaluate gradients of parameters is $N_g=2m$, where m is the number of parameterized gates in the circuit. This is discussed further in Section \ref{sec:discussion}. Error bars are the standard error on the mean, $\sigma_f /\sqrt{n}$ where $n$ is the number of examples and $\sigma_f$ the standard deviation of the performance of the optimizers. The oscillatory behaviour shown by evolutionary strategies is a feature of the algorithm, and discussed in Section~\ref{sec:discussion}. Note that negative values of $\mathcal{G}$ are observed, corresponding to on average performing worse than the initial evaluation.}
    \label{fig:results}
\end{figure*}

\subsection{Training the meta-learner}

% How was it used across all classes
For the MAX-2-SAT and Graph Bisection problems the model was trained on just 200 problems, whereas in the case of optimizing Fermi-Hubbard models the meta-learning model quickly converged and training was truncated at 100 problems. The loss function is given in Equation~(\ref{eq:loss}), where values $\omega_t = 1 \,\, \forall \,\, t$ are used. For the preprocessing of the gradients, the hyperparameter $r$ in Equation~(\ref{eq:preprocessing}) is set to $10$. For all training an Adam optimizer \cite{kingma2014adam} was used with a learning rate of $0.003$, $\beta_0=0.9$, $\beta_1=0.999$, $\epsilon=1.0^{-8}$ and zero weight decay. These training schedules were consistent across simulation type (\textit{Wave Function}, \textit{Sampling} and \textit{Noisy}). 
We included a `curriculum' method, implemented in \cite{chen2017learning}, whereby the time-horizon of the meta-learner is extended slowly throughout the training cycle. This was started at 3 iterations and capped at 10, at the end of the training cycle. Optimization was terminated if it converged before the 100 iterations, under standard convergence criteria. Overall, 9 models were trained (3 simulation environments x 3 problem classes). 
\begin{figure}
    \centering
    \includegraphics[width=0.5\textwidth]{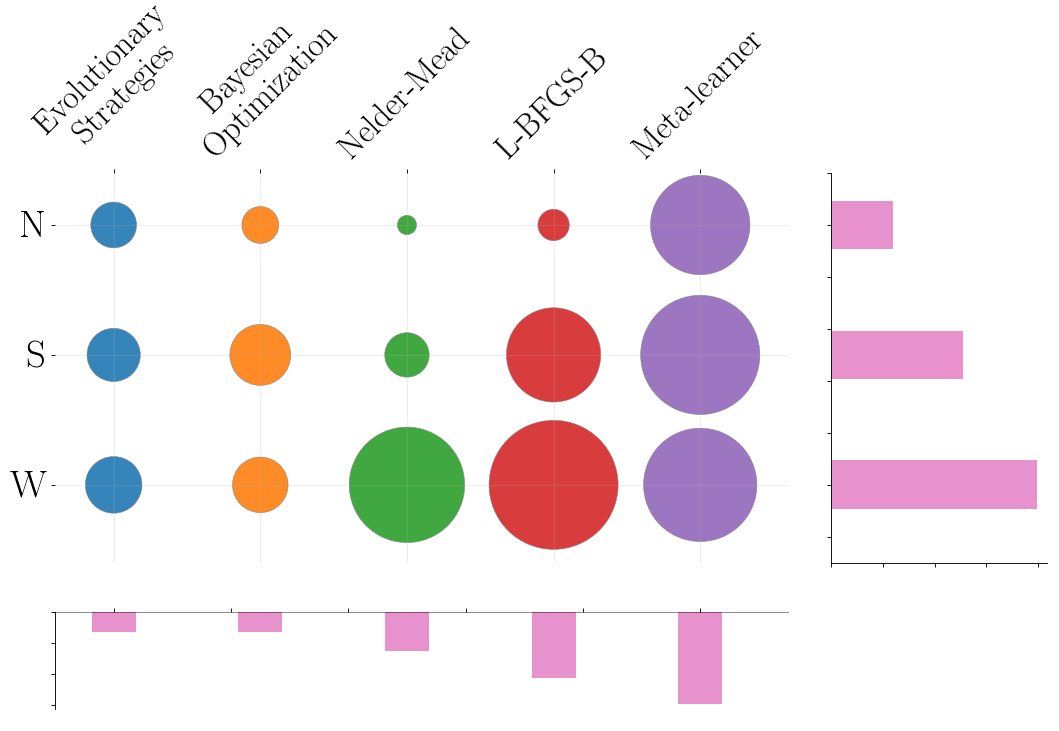}
    \caption{Bubble and bar plots of the frequency of near-optimal solutions. The size of each bubble is dependent on the total number of times an optimizer came within $2\%$ of the global optima across all problem instances (computed by Equation~(\ref{eq:D})); the largest bubble is L-BFGS-B in the \textit{Wave Function} environment (115). Repetitions are included, i.e. if an optimization ended in a near-optimal solution it was counted, regardless of whether it was found in a previous optimization. We found that if one optimizer performed well in one task, it performed well, relative to the other optimizers, in another (by this metric), so each bubble is not divided into each problem class. The right bar plot represents the summation across optimizers within a simulation type. The bottom bar plot represents the summation within an optimizer across simulation types. (N - \textit{Noisy}, S - \textit{Sampling}, W - \textit{Wave Function})}
    \label{fig:bubble}
\end{figure}

\begin{figure}
\centering
\includegraphics[width=0.48\textwidth]{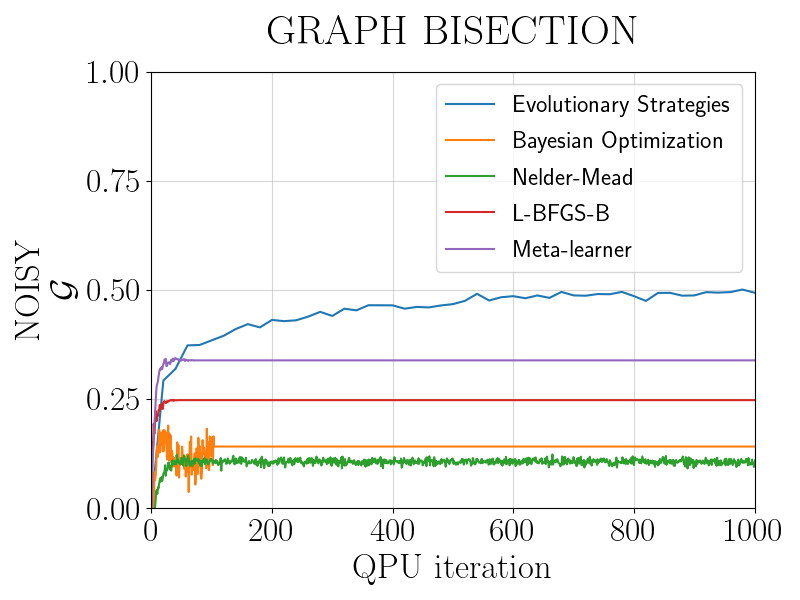}
\caption{This plot shows $\mathcal{G}$, Equation~(\ref{eq:G}), over the long timescale. This graph contains the same data as the subplot Graph Bisection, \textit{Noisy}, in Figure~\ref{fig:long_time}. The only difference is for evolutionary strategies: Only the best individuals from each generation were plotted (i.e. the individuals every 20 iterations). In the cases that the optimization was terminated at QPU iteration $< 1000$, the final value of $\mathcal{G}$ was used to extend the data.}
\label{fig:long_time}
\end{figure}

\section{Discussion \& Results}\label{sec:discussion}
Figure~\ref{fig:results} shows the performance of the optimizers measured by the gain metric in the three simulation environments. The gain metric converges in the same sense as an optimizer converging on one problem instance, this is as expected given it is an average over many problem instances. A value close to $1$ is desirable, indicating the ability of an optimizer to progress to the global minima from a starting point. Figure~\ref{fig:bubble} shows the total number of near-optimal solutions found by each optimizer. We define near-optimal as finding a solution within $2\%$ of the global optima computed by Equation~(\ref{eq:D}). The closest comparable competitor to the meta-learner in these plots is L-BFGS-B, given both optimizers had access to the gradients. This is reflected in their performance, particularly in Figure~\ref{fig:results}.

It is important to recognize that the comparison in Figure \ref{fig:results} has limited scope: The x-axis (QPU iteration) represents evaluations of $\Braket{H}$. The gradient-based optimizers (meta-learner and L-BFGS-B) evaluate auxiliary quantum circuits many times in order to compute the gradients. In order to highlight this, we have plotted the case of the worst performing meta-learner (Graph Bisection - \textit{Noisy}) in Figure \ref{fig:long_time}, where evolutionary strategies outperforms over the long timescale, though most optimizers are heavily damped by the parameter setting noise. Recognizing there are always limitations to comparing optimization methods, we draw conservative conclusions.

\subsection{General Performance}

Additionally to meta-learning functioning as an optimizer in variational quantum algorithms, we find competitive performance of this meta-learning algorithm, at small instance size, over a range of problem classes, using the gain metric $\mathcal{G}$ defined in Equation~(\ref{eq:G}); see Figure~\ref{fig:results}.

%%% Metric discussion
The metric $\mathcal{G}$ was used to evaluate and compare the optimizers, though this value can hide significant features. For example, an optimizer that finds good (but not optimal) solutions frequently will perform better than an optimizer that finds bad solutions frequently and optimal solutions infrequently. There are other cases that the reader may have in mind. 
This particular example is addressed in Figure~\ref{fig:bubble}. The number of times the optimizer comes within $2\%$ of the ground-state (across all problems), as calculated by Equation~(\ref{eq:D}), is counted. We observe an expected reduction in performance as noise is increased; this is discussed further in the subsection below. 

\subsection{Noise}
As expected, there is a reduction in performance for all optimizers as `noise' increases: Performance is worse in \textit{Sampling} than in \textit{Wave Function} and is worse in \textit{Noisy} than in \textit{Sampling}. What is notable is that the meta-learner is more resilient to this increase in noise than other methods. For example, in Fermi-Hubbard model problems, L-BFGS-B performance reduces by $0.35$ whereas the meta-learner only reduces by $0.2$, from around the same starting point (Fermi-Hubbard models column, Figure \ref{fig:results}). This pattern is repeated across problem classes, to varying degrees. We believe this is a promising sign that meta-learning will be especially useful in noisy near-term quantum heuristics implemented on hardware. In the case of simulation, we believe this resistance can be explained by the optimizer knowing how to find generally good parameters, having learned from noisy systems already. This needs to be distinguished from another potential benefit of these algorithms, where the models learn how to optimize in the presence of hardware-specific traits. In the latter case, the meta-learner may learn a model that accounts for hardware specific noise.
Further, in Figure~\ref{fig:bubble} we see a reduction in performance, measured by the total number of near-optimal solutions, for all optimizers. However, this effect is least apparent in the global optimizers (Bayesian optimization and evolutionary strategies) and the meta-learner. Additionally, the meta-learner finds significantly more near-optimal solutions (80) for \textit{Noisy} simulation than the next best optimizer (evolutionary strategies - 17). These are promising results on the potential use cases of these optimizers in hybrid algorithms implemented on noisy quantum hardware. 

\subsection{Evolutionary Strategies}
Evolutionary strategies exhibit an oscillatory behavior when gain to global optima versus function call is plotted as shown in Figure \ref{fig:results}. The first generation corresponds to a random search, then the fittest individual (i.e. best solution) found in the previous generation is evaluated first in the next generation. Hence, we observe a spike in performance every 21 evaluations (the size of the population plus the fittest individual). As we plot calls to the QPU on the x-axes of Figure \ref{fig:results}, the performance of evolutionary strategies are inaccurately represented. We plot a long-time optimization in the worst case for the meta-learner (Graph Bisection - \textit{Noisy}). In this case, we see $\mathcal{G}$ tend to a significantly higher value. Other analysis including different comparison metrics will be needed to determine the respective use cases of meta-learners vs evolutionary strategies. Indeed, while Figure \ref{fig:long_time} suggests that evolutionary strategies perform well for particularly hard problems, preliminary results in Figure \ref{fig:bubble} indicate that the meta-learner tends to outperform evolutionary strategies when searching for a near-optimal solution.

\subsection{Problems and algorithms}
The Fermi-Hubbard models were the simplest to solve (they are small problems confined to parameter values [-2,2]). This is reflected in the performance of the gradient-based optimizers. The global optimizers underperform. This is most likely a result of the size of the parameter space: Though the problem size (in terms of the number of variables) is smaller, there are significantly more parameters in this implementations we have considered of \gls{vqe}~(24) than \gls{qaoa}~(6). 

%We also see that 
Of the two classical optimization problems we consider, 
the Graph Bisection problem %explored here are
is harder than MAX-2-SAT, in the sense of worse 
classical approximability. While MAX-2-SAT can be approximated up to a constant factor, the best classical efficient algorithms known for Graph Bisection perform worse with increasing problem size~\cite{christos1994papadimitriou,ausiello2012complexity}.
This contrast appears in the performance of all optimizers: In general, every optimizer performs worse in Graph Bisection than in MAX-2-SAT by the gain metric. Bayesian optimization is the exception, showing significantly more robustness. 

\section{Conclusion}\label{sec:conc}
In this work we compared the performance of a range of optimizers (L-BFGS-B, Nelder-Mead, Bayesian optimization, evolutionary strategies and a meta-learner) across problem classes (MAX-2-SAT, Graph Bisection and Fermi-Hubbard Models) of quantum heuristics (\gls{qaoa} and \gls{vqe}) in three simulation environments (\textit{Wave Function}, \textit{Sampling} and \textit{Noisy}). We highlight two observations. The first is that the meta-learner outperforms L-BFGS-B (the closest comparable competitor) in most cases, when measured by an average percent gain metric~$\mathcal{G}$. Secondly, the meta-learner performs better than all optimizers in the \textit{Noisy} environment, measured by a total number of near-optimal solutions metric~$\mathcal{D}$. We conclude that these are promising results for the future applications of these tools to optimizing quantum heuristics, because these tools need to be robust to noise and we are often looking for near-optimal solutions. 

During the production of this work a related preprint \cite{verdon2019learning} was posted online.
In that preprint, the authors consider only gradient-free implementations of meta-learners. Their training set is orders of magnitude larger, as the meta-learner is learning to optimize from more limited information. They make similar conclusions regarding the potential of these methods and suggest using them as an initialization strategy. We broadly agree with the conclusions reached therein. 

The meta-learning methods evaluated here are relatively new and are expected to continue to improve in design and performance \cite{wichrowska2017learned}. There are several paths forward, we highlight some here. Though there is no investigation into the scaling of meta-learner performance to larger problem sizes, this in part is limited by the inability to simulate large quantum systems quickly, and exacerbated by the further burden of computing the gradients. It is an open question as to how meta-learners will perform with quantum heuristics applied to larger problem sizes. In a closely related vein, these methods will be explored on hardware implementations, for two reasons. 
The first is that quantum computing will soon be %out of 
beyond the realm of reasonable simulation times, and testing these algorithms on systems with higher number of variables will have to be done on hardware. The second is that these 
meta-learners may be able to learn hardware-specific features. For example, in this work the meta-learner is a single model applied to different parameters. This approach is called `coordinatewise'. If instead applied in a `qubitwise' fashion, where different models are trained for parameters corresponding to each qubit in a given hardware graph, there may be local variability in the physics 
%physical environment 
of each qubit that the meta-learner accounts for in its model and optimization.

In terms of further investigations into the specifics of the problems and quantum heuristics considered, we emphasize that our QAOA implementation of Graph Bisection used a different type of mixer and initial state than MAX-2-SAT.
An important question to answer is to what degree the %reduction 
differences
in performance we observed between MAX-2-SAT and Graph Bisection are due to the change of mixer and initial state, as opposed to the change of problem structure. Additional possible mixer variants and initial states for Graph Bisection are suggested in \cite{hadfield2019quantum}, which we expect to further affect QAOA performance, and hence also affect the performance of our parameter optimization approaches. An important open area of research is to better characterize the relative power of different QAOA mixers and the inherent tradeoffs in terms of performance, resource requirements, and the difficulty of finding good algorithm parameters.
In this direction, recent work \cite{wang2019xy} has demonstrated that superposition states may perform better than computational basis states as QAOA initial states.

Finally, heuristics play a prominent role in solving real-world problems: They provide practical solutions - not necessarily optimal - for complex problems (where an optimal solution is prohibitively expensive), with reasonable amount of resources (time, memory etc.). Therefore, we see significant potential for applications of quantum heuristics, implemented not only on near-term quantum devices - especially for variational quantum algorithms - but also for hybrid computing in fault-tolerant architectures. Thus it is imperative to characterize the classical components, such as the meta-learner, that learn properties of quantum devices towards the deployment of effective quantum heuristics for important practical applications.

% Finally, heuristics play a prominent role in solving real-world problems: They are practical solutions that are not necessarily optimal. Heuristics are relevant for solving problems where finding an optimal solution is prohibitively expensive, though finding better solutions provides meaningful reward. There is great potential for  quantum heuristics implemented on near-term quantum devices, such as variational quantum algorithms, to demonstrate the first useful application of quantum computation. In hybrid schemes, it is important to find the classical components, such as the meta-learner, which work best with these near-term quantum devices.  

\subsection*{Acknowledgements}
%Thanks to
The authors thank Alejandro Perdomo-Ortiz, Thomas Vandal and Walter Vinci for useful discussions. 
We are grateful for support from NASA Ames Research Center, NASA Advanced Exploration systems (AES) program, and NASA Earth Science Technology Office (ESTO). We also appreciate support from the AFRL Information Directorate under grant F4HBKC4162G001 and the Office of the Director of National Intelligence (ODNI) and the Intelligence Advanced Research Projects Activity (IARPA), via IAA 145483. S.S., F.W., and S.H. were supported by NASA Academic Mission Services, contract number NNA16BD14C. The views and conclusions contained herein are those of the authors and should not be interpreted as necessarily representing the official policies or endorsements, either expressed or implied, of ODNI, IARPA, AFRL, or the U.S. Government. The U.S. Government is authorized to reproduce and distribute reprints for Governmental purpose notwithstanding any copyright annotation thereon.

\bibliography{bibliography}
\bibliographystyle{ieeetr}

\end{document}